\documentclass[aps,twocolumn,showpacs,superscriptaddress,
preprintnumbers]{revtex4}
\usepackage{amsfonts}
\usepackage{amsmath}
\usepackage{hyperref}
\usepackage{scalefnt}

\newtheorem{theorem}{Theorem}

\newtheorem{lemma}[theorem]{Lemma}

\newenvironment{proof}[1][Proof]{\noindent\textbf{#1.} }{\ \rule{0.5em}{0.5em}}

\DeclareMathOperator{\Tr}{Tr}
\input{tcilatex}

\begin{document}

\title{Entangling Power of Permutations}
\author{Lieven Clarisse}
\email{lc181@york.ac.uk}
\affiliation{Dept. of Mathematics, The University of York, Heslington, York YO10 5DD, U.K.}
\author{Sibasish Ghosh}
\email{sibasish@cs.york.ac.uk}
\affiliation{Dept. of Computer Science, The University of York, Heslington, York YO10
5DD, U.K.}
\author{Simone Severini}
\email{ss54@york.ac.uk}
\affiliation{Dept. of Mathematics, The University of York, Heslington, York YO10 5DD, U.K.}
\affiliation{Dept. of Computer Science, The University of York, Heslington, York YO10
5DD, U.K.}
\author{Anthony Sudbery}
\email{as2@york.ac.uk}
\affiliation{Dept. of Mathematics, The University of York, Heslington, York YO10 5DD, U.K.}

\begin{abstract}
The notion of \emph{entangling power} of unitary matrices was introduced by
Zanardi, Zalka and Faoro [PRA, 62, 030301]. We study the entangling power of
permutations, given in terms of a combinatorial formula. We show that the
permutation matrices with zero entangling power are, up to local unitaries,
the identity and the swap. We construct the permutations with the minimum
nonzero entangling power for every dimension. With the use of orthogonal
latin squares, we construct the permutations with the maximum entangling
power for every dimension. Moreover, we show that the value obtained is
maximum over all unitaries of the same dimension, with possible exception
for $36$. Our result enables us to construct generic examples of $4$-qudits
maximally entangled states for all dimensions except for $2$ and $6$. We
numerically classify,~according to their entangling power, the permutation
matrices of dimension $4$ and $9$, and we give some estimates for higher
dimensions.
\end{abstract}

\maketitle

\section{Introduction}

The notion of entangling power of a quantum evolution was introduced by
Zanardi, Zalka and Faoro \cite{z} (see also \cite{z0, z1, z2}). Let $%
\mathcal{H}_{A}$, $\mathcal{H}_{B}$ and $\mathcal{H}=\mathcal{H}_{A}\otimes 
\mathcal{H}_{B}$ be Hilbert spaces where $\dim \mathcal{H}_{A}=\dim \mathcal{%
H}_{B}=d$. The entangling power of a unitary $U\in \mathcal{U(}\mathcal{H}%
)\cong U(d^{2})$ is the average amount of entanglement produced by $U$
acting on a given (uncorrelated) distribution of product states. As the pure
entanglement measure we use the linear entropy $S_{L}(\cdot )$ of the
reduced density matrix. For $|\psi \rangle \in \mathcal{H}$, let \cite{note1}%
\begin{equation*}
S_{L}(|\psi \rangle ):=\displaystyle{\frac{d}{d-1}(1-\Tr\rho ^{2})},\quad %
\mbox{where}\quad \rho =\Tr_{B}|\psi \rangle \langle \psi |.
\end{equation*}%
The \emph{entangling power} of $U$ is defined as \cite{z} 
\begin{equation}
\epsilon (U):=\int_{\langle \psi _{1}|\psi _{1}\rangle =1}\int_{\langle \psi
_{2}|\psi _{2}\rangle =1}S_{L}(U|\psi _{1}\rangle |\psi _{2}\rangle )d\psi
_{1}d\psi _{2},  \label{epu}
\end{equation}%
where $d\psi _{1}$ and $d\psi _{2}$ are normalized probability measures on
unit spheres. As the linear entropy is a concave function of the reduced
density matrix, it is an entanglement monotone \cite{vidal}, and therefore a
legitimate pure state entanglement measure.

Consider now the scenario where Alice and Bob share an unknown product state
and want to use a unitary operator to create a state which is as highly
entangled as possible. This means that Alice and Bob are looking for
unitaries with the maximum entangling power. Zanardi \cite{z0} observed that
there are permutation matrices with the maximum entangling power over all
unitaries in $\mathcal{U(}\mathcal{H})$ when $d$ is odd or $d=4n$, but the
case $d=4n+2$ was left open. In this paper we extend this result and study
the entangling power $\epsilon (P)$ of a permutation matrix $P$.

The paper is organized as follows. In Section II we give a combinatorial
expression for $\epsilon (P)$ (Theorem 2). In Section III we determine the
non-entangling permutations (Theorem \ref{ent}). In Section IV we construct
permutations with the maximum entangling power over all unitaries in $%
\mathcal{U(}\mathcal{H})$ for every $d\neq 6$. Moreover, we construct
permutations with the minimum entangling power for all $d$ (Theorem \ref{mib}%
). As a corollary, we give an upper bound to the number of permutations with
different entangling power. In Section V, we give a complete numerical
classification of permutation matrices of dimension $4$ and $9$ according to
their entangling power. For higher dimensions, we report numerical
estimates. Finally, in Section VI we draw conclusions and propose some open
problems.

\section{Entangling power of permutations}

For our purposes we find it useful to rewrite Equation \eqref{epu} in
another more concrete form as described by Zanardi \cite{z0} (see Equation %
\eqref{newe} in Lemma \eqref{zana}, below). The new expression for $\epsilon
(U)$ is based on a correspondence between quantum operations on bipartite
systems and certain quantum states which we are now going to recall.

Let $|\psi \rangle =\sum_{i,j}X_{ij}|ij\rangle $ be a state in $\mathcal{H}$
and let $X$ be the corresponding operator on $\mathcal{H}$, so that $%
X_{ij}=d\langle i|X|j\rangle $. The singular values of $X$ are equal to the
Schmidt coefficients of $|\psi \rangle $. The state $|\psi \rangle $ can be
written in terms of this operator as $|\psi \rangle =X\otimes I|\psi
_{+}\rangle $, where $|\psi _{+}\rangle =\frac{1}{\sqrt{d}}%
\sum_{i=1}^{d}|ii\rangle $ is a maximally entangled state and $I$ is the
identity operator. This relation establishes a bijection between pure states
in $\mathcal{H}\cong \mathcal{H}_{A}\otimes \mathcal{H}_{B}$ and operators $%
X $ acting on $\mathcal{H}_{A}\cong \mathcal{H}_{B}$.

Now, following Cirac \emph{et al.} \cite{c}, suppose $\mathcal{H}_{A}$ and $%
\mathcal{H}_{B}$ are themeselves bipartite: $\mathcal{H}_{A}=\mathcal{H}%
_{1}\otimes \mathcal{H}_{2}$ and $\mathcal{H}_{B}=\mathcal{H}_{3}\otimes 
\mathcal{H}_{4}$. We can repartition $\mathcal{H}\cong \mathcal{H}%
_{A}\otimes \mathcal{H}_{B}$ and regard it as the tensor product of $%
\mathcal{H}_{1}\otimes \mathcal{H}_{3}$ and $\mathcal{H}_{2}\otimes \mathcal{%
H}_{4}$. Applying the above construction to an operator $X_{13}$ yields a
state 
\begin{equation}
|X\rangle _{A|B}=|X\rangle _{12|34}:=\frac{1}{N}(X_{13}\otimes I_{24})|\Psi
_{+}\rangle _{13|24},  \label{vecfrommat}
\end{equation}%
where 
\begin{equation*}
|\Psi _{+}\rangle _{13|24}=\frac{1}{d}\sum_{i,j=1}^{d}|ij\rangle
_{13}\otimes |ij\rangle _{24},
\end{equation*}%
and $N$ is an appropriate normalization factor. This has an important
physical interpretation: the operator $X_{13}$, acting on the Hilbert space $%
\mathcal{H}_{A1}\otimes \mathcal{H}_{B3}$, corresponds to a quantum
operation. If this quantum operation is a tensor product of two quantum
operations, one acting on the system $1$ and another on the system $3$, then 
$|X\rangle _{12|34}$ is a product state with respect to the systems $A$ and $%
B$. In fact, $X_{13}\otimes I_{24}$ and $|X\rangle _{12|34}$ have the same
Schmidt decomposition if $X_{13}$ is a unitary operator \cite{z0}. It turns
out that if both parties $A$ and $B$ share an entangled state $|X\rangle
_{12|34}$ then the non-local quantum operation $X_{13}\otimes I_{24}$ can be
implemented with certainty by making use of local operations and classical
communication only (see Equation 6 of \cite{c}). With this in mind, the
entangling power of a unitary $U$ is related to the entanglement of $%
|U\rangle $ as defined in Equation \eqref{vecfrommat}. We use the linear
entropy as an entanglement measure. It might clarify matters to give the
following two extremal examples:

\begin{itemize}
\item (Identity) If $U=I$ then the corresponding state is given by 
\begin{equation*}
| I\rangle _{12|34}=\frac{1}{d}\sum_{i,j=1}^{d}|ii\rangle _{12}\otimes |
jj\rangle _{34},
\end{equation*}%
which is separable with respect to the split $A|B=12|34$. In this case we
have $S_{L}(|I\rangle )=0$.

\item (Swap) If $U=S=\sum_{ij}^{d}|ij\rangle \langle ji|$ then the
corresponding state is given by 
\begin{equation*}
| S\rangle _{12|34}=|\Psi _{+}\rangle _{A|B}=\frac{1}{d}\sum_{i,j=1}^{d}|ij%
\rangle _{12}\otimes |ij\rangle _{34},
\end{equation*}%
which is a maximally entangled vector. In this case we have $S_{L}(|S\rangle
)=1$.
\end{itemize}

The swap corresponds to the parties interchanging their systems, and might
therefore be regarded as the most non-local operation possible.
Nevertheless, the swap operation is non-entangling (that is, it never
creates entanglement). This discrepancy between \emph{non-local} and \emph{%
non-entangling} means that we cannot \emph{just} use the entanglement of $|
U\rangle $ as a measure for the entangling power of the matrix $U$. However
with a small modification one can still express $\epsilon (U)$ in terms of
the linear entropy, as is done in the following lemma.

\begin{lemma}[Zanardi \protect\cite{z0}]
\label{zana} The entangling power of a unitary $U\in \mathcal{U(}\mathcal{H}%
) $ is given by 
\begin{equation}
\epsilon (U)=\frac{d}{d+1}[S_{L}(|U\rangle )+S_{L}(|US\rangle
)-S_{L}(|S\rangle )],  \label{newe}
\end{equation}%
where $0\leq \epsilon (U)\leq \frac{d}{d+1}$ \cite{note2}.
\end{lemma}

A \emph{permutation} of $[n]=\{1,2,...,n\}$ is a bijection from $[n]$ to
itself. Every permutation $p$ of $[n]$ induces an $n\times n$ matrix $%
P=(p_{ij})$, called a \emph{permutation matrix}, such that $p_{ij}=1$ if $%
p(i)=j$ and $p_{ij}=0$ otherwise. Equivalently a permutation on $[n]$
induces a linear map of an $n$-dimensional Hilbert space which permutes a
given basis of the space (a \emph{permutation operator} on $\mathcal{H}$).
If $n=d^{2}$ we can replace $[n]$ by $[d]\times \lbrack d]$ and write $%
p(i,j)=(k_{ij},l_{ij})$; thus a permutation of $[d^{2}]$ is represented by a
pair of $d\times d$ matrices $K=(k_{ij})$ and $L=(l_{ij})$. The
corresponding permutation operator permutes the elements $|i\rangle |
j\rangle $ of a product basis of $\mathcal{H}$:%
\begin{equation*}
P(|i\rangle |j\rangle )=|k_{ij}\rangle |l_{ij}\rangle .
\end{equation*}

\begin{theorem}
Let $P=\sum_{ij}|k_{ij}l_{ij}\rangle \langle ij|$\ be a permutation matrix
in $\mathcal{U}(\mathcal{H})$. The entangling power of $P$\ is given by%
\begin{equation*}
\epsilon (P)=\frac{d^{4}+d^{2}-Q_{P}-Q_{PS}}{d(d-1)(d+1)^{2}},
\end{equation*}%
with 
\begin{equation}
Q_{P}=\sum_{i,j,m,n=1}^{d}a_{ijm}a_{ijn}b_{imn}b_{jmn},  \label{defQ}
\end{equation}%
where 
\begin{eqnarray*}
a_{ijm} &=&\langle l_{im}|l_{jm}\rangle =a_{jim}, \\
b_{imn} &=&\langle k_{im}|k_{in}\rangle =b_{inm}.
\end{eqnarray*}%
The quantity $Q_{PS}$\ is the corresponding expression for the permutation $%
PS$.
\end{theorem}

\begin{proof}
> From Lemma \ref{zana}, 
\begin{equation*}
\epsilon (P)=\frac{d^{4}+d^{2}-d^{4}[\text{Tr}\rho _{PS}^{2}+\text{Tr}\rho
_{P}^{2}]}{d(d-1)(d+1)^{2}},
\end{equation*}%
where $\rho _{P}$ and $\rho _{PS}$ are the reduced density matrices of the
states $|P\rangle \langle P|$ and $|PS\rangle \langle PS|$, respectively. We
can express $d^{4}$Tr$\rho _{P}^{2}$ in terms of the matrices $(k_{ij})$ and 
$(l_{ij})$. Applying the formula in Equation \ref{vecfrommat}, we find the
state corresponding to $P$ as 
\begin{equation*}
| P\rangle _{12|34}=\frac{1}{d}\sum_{i,m=1}^{d}|k_{im}i\rangle _{12}\otimes
| l_{im}m\rangle _{34}.
\end{equation*}%
This leads to 
\begin{equation*}
\rho _{P}=\text{Tr}_{B}|P\rangle \langle P|=\frac{1}{d^{2}}%
\sum_{i,j,m=1}^{d}a_{ijm}|k_{im}i\rangle \langle k_{jm}j|.
\end{equation*}%
By squaring and taking the trace, we obtain 
\begin{equation*}
d^{4}\text{Tr}\rho _{P}^{2}=\sum_{i,j,m,n=1}^{d}a_{ijm}a_{ijn}b_{imn}b_{jmn}.
\end{equation*}%
The same analysis applies to $Q_{PS}$.
\end{proof}

\bigskip

QTP{ite} The problem of classifying bipartite permutation operators
according to their entangling power now reduces to finding what different
values can be taken by $Q_{P}+Q_{PS}$. Observe that the coefficients $%
a_{ijn} $ and $b_{imn} $ (which are either $0$ or $1$), and the combination 
\begin{equation*}
r_{ijmn}=a_{ijm}a_{ijn}b_{imn}b_{jmn},
\end{equation*}%
which occurs in \eqref{defQ}, have the following interpretation:

\begin{itemize}
\item $a_{ijn} = 1$ if and only if $(i,n)$ and $(j,n)$, which are in the
same column of the square $[d]\times[d]$, are taken by $P$ to elements in
the same column;

\item $b_{imn} = 1$ if and only if $(i,m)$ and $(i,n)$, which are in the
same row of the square, are taken by $P$ to elements in the same row;

\item $r_{ijmn}=1$ if and only if $(i,m),(i,n),(j,m)$ and $(j,n)$, which
form a rectangle in $[d]\times \lbrack d]$, are taken by $P$ to a rectangle
with the same orientation.
\end{itemize}

We will denote the rectangle $(i,m),(i,n),(j,m),(j,n)$ (formed by the
intersection of rows $i$ and $j$ and columns $m$ and $n$) as $R_{ijmn}$. We
note that if this rectangle is non-degenerate, that is $i\neq j$ and $m\neq
n $, then it contributes either $0$ or $4$ to $Q_{P}$, since 
\begin{equation*}
r_{ijmn}=r_{jimn}=r_{ijnm}=r_{jinm}=0\text{ or }1;
\end{equation*}%
if $i=j$ or $m=n$ but not both, $R_{ijmn}$ contributes $0$ or $2$; while if $%
i=j$ and $m=n$, then $R_{ijmn}$ contributes $1$.

QTP{ite} 

QTP{ite} 

\section{Non-entangling permutations}

QTP{ite} Two permutation matrices $P,Q\in \mathcal{U(H)}$ are said to be 
\emph{locally unitarily connected} (for short, \emph{LU-connected}) if there
are unitaries $V$ acting on $\mathcal{H}_{A}$ and $W$ on $\mathcal{H}_{B}$
such that $(V\otimes W)P=Q$. Then $V$ and $W$ are actually permutation
operators. Note that if two permutations are LU-connected then they have the
same entangling power. The set of non-entangling permutation matrices is
denoted by $E^{0}$. The following result seems to be known to specialists
(see, \emph{e.g.}, \cite{rez}), but we provide a proof for completeness

\begin{theorem}
\label{ent}Let $P\in\mathcal{U(H)}$ be a permutation matrix. Then $P\in
E^{0} $ if and only if one of the following two conditions is satisfied:

\begin{enumerate}
\item $P$ is LU-connected to $I$;

\item $P$ is LU-connected to $S$.
\end{enumerate}
\end{theorem}

\begin{proof}
The permutation matrix $P$ is non-entangling if $P(|\varphi \rangle |\chi
\rangle )$ is a product state for all $|\varphi \rangle \in \mathcal{H}_{A}$
and $|\chi \rangle \in \mathcal{H}_{B}$. Consider two basis elements $|
i\rangle |j_{1}\rangle $ and $|i\rangle |j_{2}\rangle $, and suppose 
\begin{align*}
P(|i\rangle |j_{1}\rangle )& =|i_{1}^{\prime }\rangle |j_{1}^{\prime
}\rangle , \\
P(|i\rangle |j_{2}\rangle )& =|i_{2}^{\prime }\rangle |j_{2}^{\prime
}\rangle .
\end{align*}%
Then either $i_{1}^{\prime }=i_{2}^{\prime }$ or $j_{1}^{\prime
}=j_{2}^{\prime }$, otherwise $P(|i\rangle (|j_{1}\rangle +|j_{2}\rangle ))$
would be entangled. Suppose $i_{1}^{\prime }=i_{2}^{\prime }=i^{\prime }$.
Then for all $j_{3}$ we must have $P(|i\rangle |j_{3}\rangle )=|i^{\prime
}\rangle |j_{3}^{\prime }\rangle $, for if not $P(|i\rangle |j_{3}\rangle
)=|i^{\prime \prime }\rangle |j_{1}^{\prime }\rangle $ and then $P(|i\rangle
(|j_{2}\rangle +|j_{3}\rangle ))$ is entangled. So there is a permutation $%
p_{B}$ such that $P(|i\rangle |j\rangle )=|i^{\prime }\rangle |
p_{B}(j)\rangle $. Thus as a permutation of the square lattice $%
\{(i,j):1\leq i,j\leq d\}$, $P$ takes the row $(i,\cdot )$ to a row.
Consider an element $(i_{2},j_{1})$ in the same column as $(i,j_{1})$. The
permutation $p$ cannot take $(i_{2},j_{1})$ to an element in the row $%
(i^{\prime },\cdot )$, because that is already full. So $p$ must take $%
(i_{2},j_{1})$ to an element in the column $(\cdot ,j_{1}^{\prime })$. Thus $%
p$ takes the column $(\cdot ,j_{1})$ to a column. It then follows that $p$
takes every row to a row and every column to a column, that is $P((|i\rangle
| j\rangle )=|p_{A}(i)\rangle |p_{B}(j)\rangle _{2}$ for some permutation $%
p_{A}$.

In the second case, $j_{1}^{\prime }=j_{2}^{\prime }$, the permutation $p$
takes two elements in a row to elements in a column, and we can similarly
show that it takes every row to a column and every column to a row, that is $%
P(|i\rangle |j\rangle )=|p_{A}(j)\rangle |p_{B}(i)\rangle $.
\end{proof}

\bigskip

QTP{ite} The probability of sampling a permutation matrix in $E^{0}$ over
all permutations of $[d^{2}]$ goes to $0$ as $d\rightarrow \infty $. In
fact, by Theorem \ref{ent}, the number of elements in $E^{0}$ is $2(d!)^{2}$%
, and therefore the probability is 
\begin{equation*}
\begin{tabular}{lll}
$\dfrac{2(d!)^{2}}{d^{2}!}\rightarrow 0$ & as & $d\rightarrow \infty $.%
\end{tabular}%
\end{equation*}

\section{Entangling permutations}

\subsection{Maximum}

QTP{ite} In this section we construct and count the permutations with the
maximum entangling power $d/(d+1)$ that can be attained by any unitary
operator on $\mathcal{H}_{A}\otimes \mathcal{H}_{B}$. We will make use of
latin squares. Recall that a \emph{latin square} of \emph{side }$d$ is a $%
d\times d$ matrix with entries from the set $[d]=\{1,\ldots ,d\}$ such that
every row and column is a permutation of $\{1,\ldots ,d\}$, and two $d\times
d$ latin squares $(k_{ij})$ and $(l_{ij})$ are \emph{orthogonal} if $%
(k_{ij},l_{ij})$ is a permutation of $[d]\times \lbrack d]$. Equivalently,
two $d\times d$ latin squares $(k_{ij})$ and $(l_{ij})$ are orthogonal if
the ordered pairs $(k_{ij},l_{ij})$ are distinct for all $i$ and $j$. Euler
believed that there are no orthogonal latin squares of side $4n+2$. Only in
1960 was it shown by Bose, Shrikhande and Parker that, except for side $6$,
Euler's conjecture was false \cite{bsp}. The study of latin squares is an
important area of combinatorics with connections to design theory,
projective geometries, graph theory, \emph{etc.} \cite{crc, lm, ls}.
Orthogonal latin squares have been recently applied in quantum information
theory \cite{ww, hhh, kg}.

\begin{theorem}
\label{max}Let $P$ be a permutation operator on $\mathcal{H}_{A}\otimes 
\mathcal{H}_{B}$ defined by $P(|i\rangle |j\rangle )=|k_{ij}\rangle |
l_{ij}\rangle $. Then the entangling power of $P$ equals the maximum value $%
\epsilon (P)=d/(d+1)$ over $\mathcal{U(H)}$ if and only if the matrices $%
(k_{ij})$ and $(l_{ij})$ are orthogonal latin squares.
\end{theorem}

\begin{proof}
By Theorem 2, the entangling power of $P$ is maximised when $Q_{P}+Q_{PS}$
is minimised. Now $Q_{P}$ is equal to the number of rectangles in $[d]\times
\lbrack d]$ which are taken to rectangles by $P$, with the horizontal lines
remaining horizontal and the vertical lines remaining vertical. This is at
least $d^{2}$, since every rectangle consisting of a single point must be
taken to a rectangle. It is precisely $d^{2}$ if and only if no nonzero
horizontal line is taken to a horizontal line and no nonzero vertical line
is taken to a vertical line, i.e. if 
\begin{equation*}
k_{im}\neq k_{in}\quad \text{ whenever }\quad m\neq n
\end{equation*}%
and 
\begin{equation*}
l_{im}\neq l_{jm}\quad \text{ whenever }\quad i\neq j.
\end{equation*}%
On the other hand, $Q_{PS}$ is equal to the number of rectangles in $%
[d]\times \lbrack d]$ which are taken to rectangles by $P$, but with the
horizontal lines becoming vertical and the vertical lines becoming
horizontal. This will be precisely $d^{2}$ if and only if no nonzero
vertical line is taken to a horizontal line and vice versa, that is if and
only if 
\begin{equation*}
k_{im}\neq k_{jm}\quad \text{ whenever }\quad i\neq j
\end{equation*}%
and 
\begin{equation*}
l_{im}\neq l_{in}\quad \text{ whenever }\quad m\neq n.
\end{equation*}%
Together, these are the conditions for the matrices $(k_{ij})$ and $(l_{ij})$
to be latin squares. Since $(k_{ij},l_{ij})$ form a permutation of $(i,j)$,
the two latin squares are orthogonal.
\end{proof}

\bigskip

\noindent \textbf{Corollary }\label{mib1} \emph{For every }$d\neq 2,6$\emph{%
\ there is a permutation matrix }$P\in \mathcal{U(H)}$\emph{\ such that }$%
\epsilon (P)$\emph{\ is maximum over }$\mathcal{U(H)}$.

\bigskip

\begin{proof}
It follows from Theorem \ref{max} together with the fact that there are two
orthogonal latin squares for every $d\neq 2,6$ (see, \emph{e.g.}, \cite{ls}).
\end{proof}

\bigskip

As an example of a permutation matrix satisfying Theorem 4, consider%
\begin{equation*}
R=\left[ 
\begin{array}{ccc|ccc|ccc}
1 & 0 & 0 & 0 & 0 & 0 & 0 & 0 & 0 \\ 
0 & 0 & 0 & 0 & 0 & 1 & 0 & 0 & 0 \\ 
0 & 0 & 0 & 0 & 0 & 0 & 0 & 1 & 0 \\ \hline
0 & 0 & 0 & 0 & 0 & 0 & 0 & 0 & 1 \\ 
0 & 1 & 0 & 0 & 0 & 0 & 0 & 0 & 0 \\ 
0 & 0 & 0 & 1 & 0 & 0 & 0 & 0 & 0 \\ \hline
0 & 0 & 0 & 0 & 1 & 0 & 0 & 0 & 0 \\ 
0 & 0 & 0 & 0 & 0 & 0 & 1 & 0 & 0 \\ 
0 & 0 & 1 & 0 & 0 & 0 & 0 & 0 & 0%
\end{array}%
\right] .
\end{equation*}%
For the permutation matrix $R$ we have $\epsilon (R)=\frac{3}{4}$ which is
the maximum over all unitaries in $\mathcal{U}(\mathcal{H})=(\mathcal{H}%
_{A}\otimes \mathcal{H}_{B})$ where $\dim \mathcal{H}_{A}=\dim \mathcal{H}%
_{B}=3$.

By looking at $d^{2}\times d^{2}$ permutation matrices as made up of $d^{2}$
blocks, we can state an alternative version of Theorem \ref{max}.

\begin{theorem}
\label{fourway}Let $P\in \mathcal{U(H)}$ be a permutation matrix. Then $%
\epsilon (P)$ is maximum over $\mathcal{U(H)}$ if and only if $P$ satisfies
the following conditions:

\begin{enumerate}
\item Every block contains one and only one nonzero element;

\item All blocks are different;

\item Nonzero elements in the same block-row are in different sub-columns;

\item Nonzero elements in the same block-column are in different sub-rows.
\end{enumerate}
\end{theorem}

\begin{proof}
By Theorem 2, the quantity $Q_{P}$ is maximum if and only if $%
a_{ijm}=b_{imn}=1$ for all $1\leq i,j,n,m\leq d$. In this case $Q_{P}=d^{4}$
(for example, when\thinspace $P=I$). On the other hand $Q_{P}$ is minimum if
and only if $a_{ijn}=\delta _{ij}$ and $b_{imn}=\delta _{mn}$, and in this
case the sum reduces to 
\begin{equation*}
\begin{tabular}{lll}
$Q_{P}=\sum_{i,n=1}^{d}a_{iin}b_{inn}=d^{2}$ &  & (for example, when $P=S$).%
\end{tabular}%
\ 
\end{equation*}%
To obtain the maximum entangling power we need to find the permutation
matrix that minimizes $Q_{P}+Q_{PS}$. We have seen that the maximum
entangling power is $d/(d+1)$, and in \cite{z0} is shown that this value can
be obtained if and only if 
\begin{equation*}
S_{L}(|U\rangle )=S_{L}(|US\rangle )=S_{L}(|S\rangle )
\end{equation*}%
or, equivalently, 
\begin{equation*}
Q_{P}=Q_{PS}=Q_{S}=d^{2}.
\end{equation*}%
It is then easy to observe that the conditions $1$ and $2$ express the
minimality of $Q_{P}$ and the conditions $3$ and $4$ express the minimality
of $Q_{PS}$.
\end{proof}

\bigskip

Theorem \ref{fourway} implies that in a $d^{2}\times d^{2}$ permutation
matrix $P$ attaining the maximum value $d/(d+1)$, every block contains one
and only one nonzero entry (as in the above permutation matrix $R$). It is
then possible to represent $P$ by a $d\times d$ array $\widetilde{P}=(%
\widetilde{p}_{ij})$. The cell $\widetilde{p}_{ij}$ specifies the
coordinates of the nonzero entry in the $ij$-th block of $P$. For the above
permutation matrix $R$, we have 
\begin{equation*}
\widetilde{R}=%
\begin{tabular}{|ccc|}
\hline
$11$ & $23$ & $32$ \\ 
$22$ & $31$ & $13$ \\ 
$33$ & $12$ & $21$ \\ \hline
\end{tabular}%
\text{ }.
\end{equation*}%
Note that the $ij$-th cell of $\widetilde{R}$ is of the form $%
(k_{ij},l_{ij}) $ where $K=(k_{ij})$ and $L=(l_{ij})$ are the orthogonal
latin squares%
\begin{equation*}
\begin{tabular}{lll}
$K=%
\begin{tabular}{|lll|}
\hline
$1$ & $2$ & $3$ \\ 
$2$ & $3$ & $1$ \\ 
$3$ & $1$ & $2$ \\ \hline
\end{tabular}%
$ & and & $L=%
\begin{tabular}{|lll|}
\hline
$1$ & $3$ & $2$ \\ 
$2$ & $1$ & $3$ \\ 
$3$ & $2$ & $1$ \\ \hline
\end{tabular}%
\text{ }.$%
\end{tabular}%
\end{equation*}%
It follows from Theorem \ref{fourway} that a permutation matrix $P$ has
maximal entangling power if and only if $\widetilde{P}$ is obtained by \emph{%
superimposing} two orthogonal latin squares.

Direct calculations give the following result.

\begin{theorem}
The following statements are true:

\begin{enumerate}
\item For $d=2$ the matrix $P=CNOT$ attains the value $\epsilon (P)=\frac{4}{%
9}$ which is maximum over all unitaries in $\mathcal{U(}\mathcal{H})$.

\item For $d=6$ the value $\epsilon (P)=\frac{628}{735}$ is maximum over all
permutations $P\in \mathcal{U(}\mathcal{H})$ and the maximizing $P$ is
associated to 
\begin{equation}
\widetilde{P}=%
\begin{tabular}{|llllll|}
\hline
$11$ & $22$ & $33$ & $44$ & $55$ & $66$ \\ 
$23$ & $14$ & $45$ & $36$ & $61$ & $52$ \\ 
$32$ & $41$ & $64$ & $53$ & $16$ & $25$ \\ 
$46$ & $35$ & $51$ & $62$ & $24$ & $13$ \\ 
$54$ & $63$ & $26$ & $15$ & $42$ & $31$ \\ 
$65$ & $56$ & $12$ & $21$ & $33$ & $44$ \\ \hline
\end{tabular}%
\text{ }.  \label{ptil}
\end{equation}
\end{enumerate}
\end{theorem}

\begin{proof}
\emph{Part 1.} It has been shown by Rezakhani \cite{rez} that for $d=2$, the
entangling power of any unitary $U\in U(4)$ is given by 
\begin{align}
\epsilon (U)& =\frac{1}{3}-\frac{1}{9}\times \{\cos (4c_{1})\cos (4c_{2})+ 
\notag \\
& \cos (4c_{1})\cos (4c_{3})+\cos (4c_{2})\cos (4c_{3}))\},
\end{align}%
where $c_{1},c_{2},c_{3}\in \mathbb{R}$ and $|c_{3}|\leq c_{2}\leq c_{1}\leq
\pi /4$. It is easy to show that $\epsilon (U)$ takes its maximum value $4/9$
when either $c_{1}=c_{2}=\pi /4,c_{3}=0$ or $c_{1}=\pi /4,c_{2}=c_{3}=0$.
Then, every permutation matrix which is LU-connected to any of the four
matrices attains this maximum value (see Table I, given in Section V):%
\begin{equation*}
\begin{tabular}{rr}
$CNOT=\left( 
\begin{array}{llll}
1 & 0 & 0 & 0 \\ 
0 & 1 & 0 & 0 \\ 
0 & 0 & 0 & 1 \\ 
0 & 0 & 1 & 0%
\end{array}%
\right) ,$ & $S\cdot CNOT,$ \\ 
&  \\ 
$M=\left( 
\begin{array}{cccc}
1 & 0 & 0 & 0 \\ 
0 & 0 & 1 & 0 \\ 
0 & 0 & 0 & 1 \\ 
0 & 1 & 0 & 0%
\end{array}%
\right) ,$ & $S\cdot M.$%
\end{tabular}%
\end{equation*}

\emph{Part 2.} The permutation associated to $\widetilde{P}$ arises from the
latin squares 
\begin{equation*}
\begin{tabular}{|llllll|}
\hline
$1$ & $2$ & $3$ & $4$ & $5$ & $6$ \\ 
$2$ & $1$ & $4$ & $3$ & $6$ & $5$ \\ 
$3$ & $4$ & $6$ & $5$ & $1$ & $2$ \\ 
$4$ & $3$ & $5$ & $6$ & $2$ & $1$ \\ 
$5$ & $6$ & $2$ & $1$ & $4$ & $3$ \\ 
$6$ & $5$ & $1$ & $2$ & $3$ & $4$ \\ \hline
\end{tabular}%
\end{equation*}%
and%
\begin{equation*}
\begin{tabular}{|llllll|}
\hline
$1$ & $2$ & $3$ & $4$ & $5$ & $6$ \\ 
$3$ & $4$ & $5$ & $6$ & $1$ & $2$ \\ 
$2$ & $1$ & $4$ & $3$ & $6$ & $5$ \\ 
$6$ & $5$ & $1$ & $2$ & $4$ & $3$ \\ 
$4$ & $3$ & $6$ & $5$ & $2$ & $1$ \\ 
$5$ & $6$ & $2$ & $1$ & $3$ & $4$ \\ \hline
\end{tabular}%
\text{ },
\end{equation*}%
which are \emph{`as close to being orthogonal as possible'} \cite{h}. For $P$%
, we have that $Q_{P}=40$ and $Q_{PS}=36$. Since there do not exist two
orthogonal latin squares of side $6$, these are the smallest values
obtainable, as it is explained by the following argument. The array%
\begin{equation*}
\widehat{P}=%
\begin{tabular}{|llllll|}
\hline
$11$ & $22$ & $3\mathbf{3}$ & $4\mathbf{4}$ & $55$ & $66$ \\ 
$24$ & $13$ & $46$ & $35$ & $62$ & $51$ \\ 
$56$ & $65$ & $12$ & $21$ & $43$ & $34$ \\ 
$63$ & $54$ & $25$ & $16$ & $31$ & $42$ \\ 
$45$ & $36$ & $61$ & $52$ & $14$ & $23$ \\ 
$32$ & $41$ & $5\mathbf{3}$ & $6\mathbf{4}$ & $26$ & $15$ \\ \hline
\end{tabular}%
\end{equation*}%
represents the action of $P$ on the set $[6]\times \lbrack 6]$. The $ij$-th
cell of $\widehat{P}$ is $kl$ if in $P$ the contribution of the term $%
|kl\rangle \langle ij|$ is nonzero (for example, the $22$-th cell of $%
\widehat{P}$ is $13$ because $|13\rangle \langle 22|$ is nonzero in $P$). We
have printed in boldface the symbols that occur twice in a row or in a
column. Observe that, given any permutation matrix $P\in U(36)$, since every
element of the set $[6]\times \lbrack 6]$ has to occur exactly once in $%
\widehat{P}$, it is never possible to have only one pair of equal symbols in
the same column, without this being the case for also another column. Hence,
the minimum value attainable by $Q_{P}$ would be $2$ steps ahead of $36$,
that is $40$.
\end{proof}

\bigskip

We have established a bijection between pairs of orthogonal latin squares of
side $d$ and $d^{2}\times d^{2}$ permutations with maximal entangling power.
The number of pairs of orthogonal latin squares of side $d$ is known only
for small $d$ (see A072377, \cite{sloan}). It is $36$ for $d=3$ and $3456$
for $d=4$. Note that these values apply to unordered pairs of orthogonal
latin squares. This means that the number of $d^{2}\times d^{2}$
permutations with the maximum entangling power is twice the number of pairs
of orthogonal latin squares of side $d$. General methods for constructing
pairs of orthogonal latin squares are presented in \cite{gras, ls, l}.

\bigskip

For a unitary $U$ that reaches the upper bound $\epsilon (U)=d/(d+1)$, we
have $S_{L}(|U\rangle )=S_{L}(|US\rangle )=1$. In other words, $|U\rangle $
is maximally entangled with respect to the bipartite splits $12|34$ and $%
23|14$. It turns out that this is also maximally entangled with respect to
the split $13|24$. The easiest way to see this is by looking at Equation \ref%
{vecfrommat} defining $|U\rangle $. The state $|U\rangle _{13|24}$ is
obtained by two parties $A=13$ and $B=24$ sharing a maximally entangled
state, by applying the unitary $U$ on Alice's system. Since this is a local
reversible transformation, the entanglement is preserved, hence the state $%
|U\rangle _{13|24}$ is maximally entangled. Thus a unitary $U$ is maximally
entangling if and only $|U\rangle $ is maximally entangled with respect to
the three possible bipartite splits \cite{Zalka}. Note that $|U\rangle $ is
also maximally entangled in all the four splits $1|234$, $2|134$, $3|124$
and $4|123$. Then $|U\rangle $ is indeed a maximally entangled state. The
construction of maximally entangling unitaries thus provides us with a
canonical way of constructing such maximally entangled $4$-qudits for all
dimensions except $d=2$ and $d=6$. It has been known for some time that this
is not possible for $d=2$, \emph{i.e.}\ $4$-qubits (see \cite{tony}),
leaving only the case $d=6$ open.

\subsection{Minimum}

In this section we construct the permutations with the minimum nonzero
entangling power that can be attained by permutation operators on $\mathcal{H%
}_{A}\otimes \mathcal{H}_{B}$. Two permutation matrices $P,Q\in \mathcal{U(H)%
}$ are said to be in the same \emph{entangling class} if $\epsilon
(P)=\epsilon (Q)$.

\begin{theorem}
\label{mib} Let $P\in \mathcal{U(H)}$ be a permutation matrix. Then $%
\epsilon (P)$ is nonzero but minimum over all permutations in $\mathcal{U(H)}
$ if%
\begin{equation*}
\widehat{P}=%
\begin{tabular}{|lllll|}
\hline
$11$ & $12$ & $\cdots $ & $\cdots $ & $1d$ \\ 
$21$ & $22$ & $\ldots $ & $\ldots $ & $2d$ \\ 
$\vdots $ & $\vdots $ &  &  & $\vdots $ \\ 
$(d-1)1$ & $(d-1)2$ & $\ldots $ & $\ldots $ & $(d-1)d$ \\ 
$d1$ & $d2$ & $\ldots $ & $dd$ & $d(d-1)$ \\ \hline
\end{tabular}%
\ .
\end{equation*}%
In such a case 
\begin{equation}
\epsilon (P)=\frac{8(d-1)}{d(d+1)^{2}}.  \label{minepsilon}
\end{equation}
\end{theorem}

\begin{proof}
Since $P$ is the permutation matrix \emph{closest} to $I$, it is clear that $%
\epsilon (P)$ is miminum. To find its value, we compare $Q_{P}$ with $Q_{I}$
where I denotes the identity permutation. The rectangles $R_{ijmn}$ which
contribute to $Q_{I}$ but not to $Q_{P}$ are those containing $(d-1,d)$ or $%
(d,d)$ or both, except for the degenerate rectangles contained in the bottom
line of the square (i.e. those with $i=j=d$). There are $(d-1)(d-2)$
non-degenerate rectangles containing $(d-1,d)$ but not $(d,d-1)$, each
contributing $4$ to $Q_{I}$, and $d-1$ degenerate vertical rectangles, with
two equal vertices at $(d,d-1)$ and two equal vertices at $(i,d-1)$ where $%
i\neq d$. Each of these contributes 2 to $Q_{I}$, so the total contribution
from rectangles containing $(d,d-1)$ but not $(d,d)$ is $4(d-1)(d-2)+2(d-1)$%
. There is an equal contribution from rectangles containing $(d,d)$ but not $%
(d,d-1)$. Finally, there are $d-1$ non-degenerate rectangles containing both 
$(d,d-1)$ and $(d,d)$, each contributing 4 to $Q_{I}$. The total
contribution to $Q_{I}$ which is not included in $Q_{P}$ is $8(d-1)^{2}$.
Since $Q_{I}=d^{4}$, we have 
\begin{equation*}
Q_{P}=d^{4}-8(d-1)^{2}.
\end{equation*}%
On the other hand, $Q_{PS}=d^{2}$ since no nonzero horizontal line is taken
to a vertical line by $P$. Hence, by Theorem 2, $\epsilon (P)$ is given by %
\eqref{minepsilon}.
\end{proof}

\bigskip

\noindent \textbf{Corollary }\emph{An upper bound to the number of different
entangling classes of permutations is given by }%
\begin{equation}
B=2+\frac{1}{2}(d^{4}-d^{2}-8(d-1)^{2}).  \label{boun}
\end{equation}

\begin{proof}
We can write $Q_{P}=\sum_{i,j,m,n=1}^{d}r_{ijmn}$. If $i\neq j$ and $m\neq n$%
, the contribution of the rectangle $R_{ijmn}$ is $0$ or $4$, which is even.
If either $i=j$ or $m=n$, the contribution of $R_{ijmn}$ is $0$ or $2$,
which is again even. If $i=j$ and $m=n$, the contribution of $R_{ijmn}$ is $%
1 $, and in total we have $d^{2}$ such terms. It follows that $Q_{P}$ is
even if and only if $d$ is even. The same analysis applies to $Q_{PS}$. Then 
$Q_{P}+Q_{PS}$ is even for all $d$.

Now we have seen that the zero entangling power corresponds to $%
Q=d^{4}+d^{2} $, and the maximum to $Q=2d^{2}$, so that $\frac{1}{2}%
(d^{4}-d^{2}+2)$ is an upper bound to the number of classes, where the
coefficient $1/2$ comes from the fact that two consecutives values of $Q_{P}$
($Q_{PS}$) differ by a multiple of $2$. We can tighten this bound and obtain
the value $(B-2)$ by making use of the fact from Theorem \ref{mib} that the
value of $Q_{P}+Q_{PS} $ is $d^{4}-8(d-1)^{2}+d^{2}$ when $\epsilon (P)$ is
nonzero but minimum. The first term $2$ on the RHS of Equation \ref{boun}
occurs because of the two classes corresponding to zero and the maximum
entangling power.
\end{proof}

\section{Numerical results}

In this section we report some numerical results. We are interested in
counting the number of different entangling classes for different
dimensions. We are also interested in the average entangling power over all
permutations of a given dimension. The results are given in the following
tables:

\begin{table}[ph]
\begin{center}
\begin{tabular}{c|c}
Entangling Power $\epsilon (P)$ & Number of elements \\ 
& in entangling class \\ \hline
0 & 8 \\ 
4/9 & 16%
\end{tabular}%
\end{center}
\caption{Classes of permutations with different entangling power and the
number of elements in each class for $d=2$.}
\end{table}
\begin{table}[h]
\begin{center}
\begin{tabular}{c|c}
Entangling Power $\epsilon (P)$ & Number of elements \\ 
& in entangling class \\ \hline
0 & 72 \\ 
1/3 & 2592 \\ 
3/8 & 864 \\ 
5/12 & 1296 \\ 
182/375 & 10368 \\ 
23/48 & 20736 \\ 
1/2 & 27432 \\ 
25/48 & 36288 \\ 
13/24 & 44064 \\ 
9/16 & 101376 \\ 
7/12 & 44712 \\ 
29/48 & 46656 \\ 
5/8 & 22464 \\ 
2/3 & 3888 \\ 
3/4 & 72%
\end{tabular}%
\end{center}
\caption{Classes of permutations with different entangling power and the
number of elements in each class for $d=3$.}
\end{table}
\begin{table}[tbp]
\begin{center}
\begin{tabular}{c|c|c}
Dimension $d$ & Number of classes & Average entangling power \\ \hline
$2$ & $2$ & $\frac{8}{27}\approx 0.29$ \\ 
$3$ & $15$ & $\frac{31}{56}\approx 0.55$ \\ 
$4$ & $\geq 65$ & $0.67\pm 0.01$ \\ 
$5$ & $\geq 190$ & $0.74\pm 0.01$%
\end{tabular}%
\end{center}
\caption{Number of classes of permutations with different entangling power
and the average entangling power as a function of the dimension $d$.}
\end{table}

\section{Conclusion and open problems}

In this paper we have studied the entangling power of permutations. We have
shown that the permutation matrices with zero entangling power are, up to
local unitaries, the identity and the swap. For all dimensions, we have
constructed the permutations with (nonzero) minimum entangling power. With
the use of orthogonal latin squares, we have constructed the permutations
with the maximum entangling power for every dimension. Moreover, we have
shown that this value is maximum over all unitaries of the same dimension,
with a possible exception for $36$. Our result enabled us to construct
generic examples of $4$-qudits maximally entangled states for all dimensions
except for $2$ and $6$. We have reported numerical results about a complete
classification of permutation matrices of dimension $4$ and $9$ according to
their entangling power.

\bigskip

We conclude with a list of open problems:

\begin{itemize}
\item Describe a general classification of the permutation matrices
according to their entangling power.

\item Give a formula for the average entangling power over all permutation
matrices of a given dimension.

\item For $d=6$, does there exist $U\in \mathcal{U(}\mathcal{H})$ such that $%
\epsilon (U)>\frac{628}{735}$?

\item The formula in Equation (\ref{newe}) only works for the linear
entropy. It would be desirable to have a similar simple formula of the
entangling power of a unitary in terms of the von Neumann entropy.

\item Study the entangling power of permutation matrices in relation to
multipartite systems. In this context, it is conceivable that the
permutation matrices with maximum entangling power are related to sets of
mutually orthogonal latin squares.
\end{itemize}

\emph{Acknowledgements.} We would like to thank S. L. Braunstein for helpful
comments, and Ch. Zalka for pointing out the connection between maximally
entangling unitaries and $4$-qudit states that are maximally entangled with
respect to all bipartite splits. LC is supported by a WW Smith Scholarship.
SG and SS are supported by EPSRC grants GR/87406 and GR/S56252, respectively.


\begin{thebibliography}{99}
\bibitem{note1} In refs. \cite{z} and \cite{z0}, $S_{L}(|\psi \rangle )$ is
taken as $S_{L}(|\psi \rangle )=1-$Tr$\rho ^{2}$, where $\rho =$ Tr$%
_{B}(|\psi \rangle \langle \psi |)$. In order to have $S_{L}(|\psi \rangle
)\in \lbrack 0,1]$, we have taken $S_{L}(|\psi \rangle )=\frac{d}{d-1}(1-$Tr$%
\rho ^{2})$.

\bibitem{note2} In ref. \cite{z}, the coefficient in front of the term $%
[S_{L}(|U\rangle )+S_{L}(|US\rangle )-S_{L}(|S\rangle )]$ is $\left( \frac{%
d-1}{d+1}\right) \frac{1}{S_{L}(|S\rangle )}$, which in our case is $\frac{d%
}{d+1}$, as we used a modified definition of $S_{L}(|\psi \rangle )$.

\bibitem{bsp} R. C. Bose, S. S. Shrikhande and E. T. Parker, Further results
on the construction of mutually orthogonal Latin squares and the falsity of
Euler's conjecture, \emph{Canad. J. Math} \textbf{12} (1960), 189--203.

\bibitem{c} J. I. Cirac, W. D\"{u}r, B. Kraus, and M. Lewenstein, Entangling
Operations and Their Implementation Using a Small Amount of Entanglement, 
\emph{Phys. Rev. Lett.} \textbf{86}, 544 (2001).

\bibitem{crc} C. J. Colbourn and J. H. Dinitz (eds.), The CRC handbook of
combinatorial designs, CRC Press Series on Discrete Mathematics and its
Applications, \emph{CRC Press, Boca Raton, FL} (1996).

\bibitem{ls} J. D\'{e}nes and A. D. Keedwell (eds.), Latin Squares: New
Developments in the Theory and Applications, \emph{North-Holland, Amsterdam}
(1991).

\bibitem{gras} M.-N. Gras, Une construction explicite de carr\'{e}s latin
orthogonaux d'ordre {$n$} pair, {$n\geq 10$}, \emph{J. Alg.} \textbf{219},
369 (1999).

\bibitem{hhh} A.Hayashi, M.Horibe and T.Hashimoto, The king's problem with
mutually unbiased bases and orthogonal Latin squares, quant-ph/0502092.

\bibitem{z2} A. Hamma and P. Zanardi, Quantum entangling power of
adiabatically connected {H}amiltonians, \emph{Phys. Rev. A} \textbf{69},
062319 (2004).

\bibitem{tony} A. Higuchi and A. Sudbery, How entangled can two couples get? 
\emph{Phys. Lett. A} \textbf{273}, 213--217 (2000).

\bibitem{h} R. Hill, A First Course in Coding Theory, \emph{Clarendon Press,
Oxford} (1986).

\bibitem{kg} A. Klappenecker and M. R\"{o}tteler, Unitary error bases:
constructions, equivalence, and applications, Applied algebra, algebraic
algorithms and error-correcting codes (Toulouse, 2003), 139--149, Lecture
Notes in Comput. Sci., 2643, \emph{Springer, Berlin} (2003).

\bibitem{lm} C. F. Laywine and G. L. Mullen, Discrete mathematics using
Latin squares, Wiley-Interscience Series in Discrete Mathematics and
Optimization, A Wiley-Interscience Publication. \emph{John Wiley \& Sons,
Inc., New York} (1998).

\bibitem{l} Z. Lie, A short disproof of Euler's conjecture concerning
orthogonal Latin squares, With editorial comment by A. D. Keedwell, \emph{%
Ars Combin.} \textbf{14} (1982), 47--55.

\bibitem{rez} A. T. Rezakhani, Characterization of two-qubit perfect
entanglers, \emph{Phys. Rev. A} \textbf{70}, 052313 (2004).

\bibitem{sloan} N. J. A. Sloane, The On-Line Encyclopedia of Integer
Sequences, published electronically at:\newline
\href{http://www.research.att.com/~njas/sequences/index.html}{%
http://www.research.att.com/\symbol{126}njas/sequences/}.

\bibitem{vidal} G. Vidal, Entanglement monotones, \emph{J. Mod. Opt.} 
\textbf{47}, 355 (2000).

\bibitem{ww} K. G. H. Vollbrecht and R. F. Werner, Why two qubits are
special, \emph{J. Math. Phys.} \textbf{41} (2000), 6772--6782.

\bibitem{z1} X. Wang and P. Zanardi, Quantum entanglement of unitary
operators on bi-partite systems, \emph{Phys. Rev. A} \textbf{66}, 044303
(2002).

\bibitem{Zalka} Ch. Zalka, \emph{Private communication}.

\bibitem{z} P. Zanardi, Ch. Zalka, and L. Faoro, Entangling power of quantum
evolutions, \emph{Phys. Rev. A} \textbf{62}, 030301 (2000).

\bibitem{z0} P. Zanardi, Entanglement of quantum evolutions, \emph{Phys.
Rev. A} \textbf{63}, 040304 (2001).
\end{thebibliography}
\end{document}